\batchmode
\documentclass[aps,prb,preprint,
groupedaddress,showpacs,floatfix]{revtex4-1}
\usepackage{graphicx}\usepackage{amssymb}\usepackage{color}
\def\figref#1{Fig.~\ref{#1}}
\def\eqnref#1{Eqn.~(\ref{#1})}

\begin{document}
\title[Hot Tungsten Wires]{Temperature dependent polarization of the thermal radiation emitted by thin, hot tungsten wires}
\author{A. F. Borghesani}
\email[]{armandofrancesco.borghesani@unipd.it}
\affiliation{CNISM Unit, Department of Physics and Astronomy \\ University of Padua, Padua, Italy}
\affiliation{Istituto Nazionale di Fisica Nucleare, Sezione di Padova\\via F. Marzolo 8, I-35131 Padua, Italy}
\author{G. Carugno}\affiliation{Istituto Nazionale di Fisica Nucleare, Sezione di Padova and \\Department of Physics and Astronomy,University of Padua \\
via F. Marzolo 8, I-35131 Padua, Italy}

\date{\today}
 
\begin{abstract}
We report measurements of the temperature $T $ 
dependence of the linear polarization $\langle P\rangle $ of the thermal radiation emitted by thin, incandescent tungsten wires.
We investigated an interval ranging from a little above room temperature up to melting, $T_m\approx 3700\,$K. These are the first measurements in such wide a range. 
We found that $\langle P\rangle $ decreases with increasing $T.$ We obtained a satisfactory agreement with the theoretical predictions based on the Kirchhoff's law by using a Drude-type formula for the optical properties of tungsten. The validity of such formula is assessed in literature for $T\leq 2400 \,$K and for wavelengths in the range from visible up to $\lambda\approx 2.6\,\mu$m. We have extended the range of validity of this formula for $T$ up to $T_m$ and for $\lambda$ up to $\approx 12\,\mu$m.
\end{abstract}
     
\maketitle

\section{Introduction}\label{sect:Intro}
The study of thermal emission by hot bodies is a very important topic because the celebrated Planck's result about the spectrum of a blackbody radiator paved the way for the development of Quantum Mechanics~\cite{planck1901}. Planck's law is independent of the characteristics of the blackbody material and it only depends on temperature $T$, thus making pyrometry a universal thermometric technique~\cite{singer2011}.

According to Planck's derivation, the blackbody emission consists of unpolarized, incoherent radiation for bodies whose size is larger than the typical thermal wavelength, $\lambda_T = hc/k_\mathrm{B}T,$ where $h,$ $c,$ and $k_\mathrm{B}$ are Planck's constant, light speed, and Boltzmann's constant, respectively. 

The modern availability of radiators of size comparable to or even smaller than 
$\lambda_{T}$ has led to the discovery that thermal radiation shows a high degree of spatial and temporal coherence in the near-field region~\cite{greffet1999,greffet2002,greffet2007}. Suitable subwavelength patterning of the properties of metallo-dielectric surfaces at nanoscale leads to coherence properties of the thermal emission of such nanoheaters, including carbon nanotubes~\cite{islam2004,aliev2008,fan2009,singer2011b}, that have great relevance in applied physics and engineering~\cite{laroche2005,celanovic2005,chan2006,Ingvarsson2007,klein2009}.

Early measurements with hot, a few $\mu$m thick, W-~\cite{ohman1961} and Ag~\cite{agdur1963} wires have shown that thermal radiation has a high degree of linear polarization, up to $\approx 30\,\%$ and more, orthogonal to the wire axis. 
The observed polarization was explained in terms of plasma oscillations of the electron gas in the metal that can scatter, absorb, and emit light.
More recently, an experimental study about the degree of linear polarization of incandescent W wires of diameter 
$5\,\mu$m to $100\,\mu$m  in the visible range has been published~\cite{njp} that confirms the early observations of polarization in excess of $20\,\%,$ directed perpendicularly to the wire's axis. Unfortunately, no attempt was done to measure the wire temperature that was estimated to be $\approx 2400\, $K. 

In those studies, the wire thickness is $r\gtrsim\lambda_T.$ Recent investigations on wires with $r\lesssim\lambda_T$  have shown that the emitted radiation is polarized along the wire axis, becoming fully polarized as $r\rightarrow 0$~\cite{Ingvarsson2007,au2008}.
The observation that standing waves of thermally generated charge oscillations in the near field occur across metallic stripes a few $\mu$m wide has led to the explanation of the increased polarization as a manifestation of charge confinement and correlated fluctuations along the long axis of the nanoheater~\cite{dewilde2006}. Surface plasmon polaritons propagate only in the direction of charge oscillations. So, charge oscillations driven by the thermal environment are affected in different ways whether they are parallel or perpendicular to the heater axis when $r$ is shrinked~\cite{au2008}. When $r< \lambda_T,$ longitudinal charge fluctuations are strongly correlated by the coupling with surface plasmons and light is polarized along the heater long axis. For heater width $r\geq \lambda_T$ transversal charge oscillations get correlated via the interaction with surface plasmons and the emitted light becomes polarized perpendicular to the heater axis. Actually, a rotation of the linear polarization of light emitted by Pt nanoheaters has been observed when their width changes from submicron- to micron size or when $T$ is changed, the crossover occuring for $r$ such that $2\pi r/\lambda_T \sim 1.5$~\cite{klein2009}. 

In these latter studies, the nature of the nanoheaters material is not really important as only the ratio $r/\lambda_T$ determines the direction of light polarization. However, the coupling with surface plasmons is ruled by the properties of the dielectric constant of the material~\cite{ruda2005}, which depends on $T,$ on the wavelength $\lambda,$ and on the nature of the metal~\cite{kittel}. In general, the optical properties of materials, including optical constants and emissivity, depend on $T$ and $\lambda.$
Actually, theoretical studies address the issue of how the optical properties of the material, not only its size, influence the features of the radiation emitted by long cylinders~\cite{Golyk:2012fk} and show that the polarization curves for W may shift by a factor of 10 when $T$ is changed from $300\,$K to $2400\,$K. 

In this paper, we report measurements of the degree of linear polarization of the light emitted by W wires heated by Joule effect in a temperature range from room- up to melting temperature in a wavelength band across the infrared and visible region. Wires of radius $r=9\,\mu$m, $25\,\mu$m, 
and $50\,\mu$m, respectively, are investigated. Their size is such that the light emitted is always  polarized perpendicular to their axis. Thus, the variation of the degree of polarization can solely be ascribed to the temperature and wavelength dependence of the optical properties of tungsten.

The paper is organized as follows: in Sect.~\ref{sect:expdet} we describe the experimental apparatus. In Sect.~\ref{sect:expres} we present the experimental data and compare them with the theoretical predictions. Finally, the conclusions are drawn in Sect.~\ref{sect:conc}.

\section{Experimental Details}\label{sect:expdet}  
The experimental apparatus consists of two independent subsystems. The first one is the mechanical and optical setup necessary to support the wires and to collect the emitted light. The second one consists of the electronics required to energize the wires and to reveal and analyze the detector signal.

\subsection{Mechanical and Optical Setup}\label{sect:mechopt}
\begin{figure}[!t]
\centering
\includegraphics{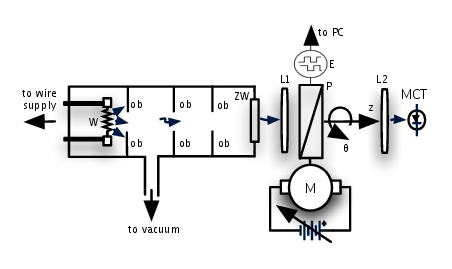}
\caption{\small Schematics of the optical setup. {\tt W} = wires, {\tt ob} = optical baffles, {\tt ZW}=~ZnSe window, {\tt L1} and {\tt L2} = lenses, {\tt P} = rotating analyzer, {\tt E} = encoder, {\tt MCT} = HgCdTe detector,  {\tt M} = motor. \label{fig:optset}}
\end{figure}

The mechanical and optical parts of the apparatus are schematically shown in \figref{fig:optset}. 
Tungsten wires ({\tt W}) of nominal purity $>99.95\,\%$, supplied by LUMA ($9\,\mu$m) and SIT ($25\,\mu$m and $50\,\mu$m), are mounted inside a $50\,$cm long metal pipe of $2.5\,$cm in diameter, evacuated to a working pressure $p\le 10^{-3}\,$Pa. The wires are stretched and clamped on supports connected to the electrical power supply by means of suitable vacuum feedthroughs. 

The wires used in this experiment are $\approx 7\,$mm long and their radius is either $50\,\mu$m, $25\,\mu$m, or $9\,\mu$m. Except for the $50\,\mu$m-sample, we used four wires at once, mounted parallel to each other, in order to increase the amount of light impinging on the detector while keeping their electrical resistance at a manageably low value. The wires are mounted with their cylindrical axes perpendicular to the axis of the vacuum pipe whose internal surface is mat and coated with Aquadag in order to minimize polarized reflections from the inner walls. Three equally spaced optical baffles ({\tt ob}) consisting of drilled washers with a central hole of $\approx 6\,$mm in diameter are located along the optical axis in order to further prevent internally reflected light from reaching the detector and to reduce the contribution of non paraxial rays. 
The light eventually exits the pipe through a ZnSe optical window ({\tt ZW}) of $\approx 1\,$cm in diameter located $\approx 30\,$cm from the wires.

Two ZnSe lenses, {\tt L1} and {\tt L2} with focal lengths of 15 and $6.5\,$ cm, respectively, image the wires on the liquid N$_{2}$ cooled, photovoltaic HgCdTe detector (Fermionics, mod. PV-12-0.5) that has a circular active area of $1\,$mm$^{2}$ and a spectral range $0.5\,\mu$m$\,\le\lambda\le 12\,\mu$m. 

The polarization degree of the light emitted by the wires is analyzed by
means of a ZnSe wire grid, infrared (IR) polarizer (Thorlabs, WP25H-Z) mounted
on a rotary frame coupled to a d.c. motor by means of a scaler gear so that it can be continuously rotated about the optical {\tt z}-axis of the system. The rotational speed is varied by changing the driving voltage of the d.c. motor. Actually, the polarizer is rotated through finite steps in order to improve the signal-to-noise ratio (SNR), as explained later. The rotation of the scaler gear shaft is measured by a 12-bit digital encoder ({\tt E}) interfaced to a PC. One complete turn of the shaft corresponds to a $2^{\circ}$ rotation of the polarizer.

A second polarizer can be inserted, if necessary, in the optical path in order to verify that no residual light outside the accessible spectral range still reaches the detector and to determine the polarization direction with respect to the wire axis orientation. It turns out that the polarization is always directed perpendicularly to that axis.

\subsection{Electronics}\label{sect:el}
In a previous experiment in the visible range~\cite{njp}, the emitted light was modulated by means of a mechanical chopper. This technique cannot be exploited in the infrared range, in which all surfaces emit detectable radiation. In this case, the tiny emitting area of the wires is negligible with respect to the much larger area of the chopper blades and of the environment that would then obscure the wire signal. 

For this reason, in the present experiment the light emission is modulated by superimposing a small, low-frequency ($\approx 2\,$Hz) a.c. current to a steady d.c. current that sets the average wire temperature. Owing to the negligible thermal inertia, the wire temperature (and emission) instantaneously follows the current changes, whereas the surrounding environment remains at constant temperature because of its enormous thermal inertia and its emission does not change as long as the d.c. current in the wires is kept constant. In this way, the a.c. wire signal is completely decoupled from the environment contribution and standard lock-in amplification techniques are used to detect it.

\begin{figure}[b!]
\centering 
\includegraphics{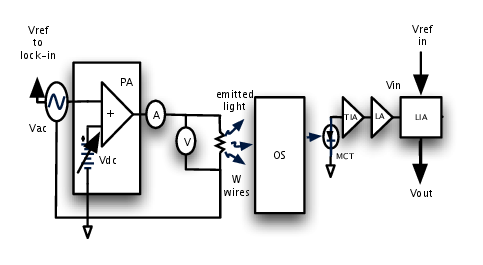}
\caption{\small Electronics of the hot wire experiment. {\tt Vac} = sinusoidal  a.c. voltage generator, {\tt Vdc} = d.c. voltage supply, {\tt PA} = power amplifier, {\tt A} = ammeter, {\tt V} = voltmeter, {\tt MCT} = HgCdTe photovoltaic detector, {\tt TIA} = transimpedance amplifier,  {\tt LA} = linear amplifier, {\tt LIA} = lock-in amplifier, {\tt OS} = optical setup. \label{fig:electronics}}
\end{figure}

The electronics required to power the wires and to measure the emitted light is shown in \figref{fig:electronics}.
A home-made, linear, modified audio power amplifier ({\tt PA}), with high-precision, adjustable internal d.c. source ({\tt Vdc}), can deliver currents up to $12\,$A from d.c. to a few tens of kHz to resistive loads with impedance $\ll 1\,\Omega$~\cite{galborg}.
The a.c. contribution is supplied by a signal generator (HP, mod. 3312A), which also issues the reference signal for lock-in detection. The output of {\tt PA} directly feeds the wires whose resistance ($0.2\,\Omega\le R\le 1\,\Omega$ at room temperature) is measured with the standard Kelvin technique by using the ammeter {\tt A} (Tektronix, mod. DMM914) and the voltmeter {\tt V} (Keithley, mod. 195A).  

The light emitted by the wires crosses the optical setup and is focused onto the photovoltaic detector {\tt MCT}. The alignment of the optical system is achieved by using a laser pointer and by finely positioning the detector with an $x-y$ translational stage. The photodiode current is converted to voltage by the transimpedance amplifier {\tt TIA} (Fermionics, PVA-500-10) whose output is linearly amplified by the amplifier {\tt LA} (EG\&G PARC, mod. 113). The output of {\tt LA} is fed to the lock-in amplifier {\tt LIA} (Stanford Research Systems, mod. SR830), whose reference signal is supplied by the signal generator {\tt Vac}. In order to maximize the LIA output, we have to manually adjust the lock-in phase reference so as to null the quadrature signal because the relative phase difference between light signal and reference voltage is unknown and because the wires act as a natural low-pass filter whose time constant is {\em a priori} unknown, too. We take advantage of the fact that the working frequency is kept constant throughout the experiment and is constantly monitored by a frequency meter (Agilent, mod. 34401A).

 \section{Experimental Results and Discussion}\label{sect:expres}

\subsection{Experimental Procedure}\label{sect:expproc}
When a good vacuum is reached in the vacuum pipe, the d.c. voltage is set to the desired working point and wires and environment are allowed to reach the working temperature. As the tungsten resistivity depends on $T$~\cite{lide1995}, the approach to equilibrium is monitored by recording the wires resistance $R.$
Once steady-state is reached, $R$ remains constant.

 It has to be noted that it is more appropriate to speak about steady-state conditions rather than equilibrium. Actually, the wires are clamped to supports that are strongly thermally coupled to the environment and the balance between Joule heating and the heat dissipation by thermal conduction over the wire boundaries and by emission of radiation leads to a strongly non uniform temperature profile along the wires. Under steady-state conditions, the temperature profile is still non uniform but does no longer change in time.
 
 When steady state is reached, the wires resistance is measured and the a.c. modulation is turned on so that the modulation signal produced by the detector can be observed and monitored. The wire resistance is also continuously monitored and recorded during the whole experimental run.

\subsection{Signal Formation and Analysis}\label{sect:signal}

The detector signal is proportional to the intensity of the light emitted by the wires, which, in turn, depends on their temperature and, thus, on 
the power dissipated into the them. As a consequence, a relationship between detector signal and electric power 
is needed.

Let us assume for a while that the wire temperature is uniform. This is not a necessary condition for the following argument and, later, this assumption will be relaxed.
Let $i$ be the current in the wires and $I$ the emitted light intensity. Let $V_{\mathrm{d.c.}}=V_{0}$ be the d.c. component of the voltage across the wires and $V_{\mathrm{a.c.}}=V_{1}\cos{\omega t} $ its a.c. component. The total current in the wires is thus
\begin{equation}
i=i_{0}+ i_{1}\cos{\omega t}
\label{eq:i}\end{equation}
in which $i_{0}=V_{0}/R$ and $i_{1}=V_{1}/R.$
The modulation amplitude is always smaller than the strength of the d.c. component. In the worst case, at low temperature, when the light emission is very weak (in this case, the glow of the wires cannot even be seen by naked eyes), $i_{1}/i_{0}= V_{1}/V_{0}\leq 0.1.$ Usually, $ V_{1}/V_{0}\approx 5\times 10^{-2}$ or less. 

Owing to the small modulation amplitude, we can assume that the wire temperature is determined, at least to first order, by the d.c. bias. 
Hence, also the wire resistance $R$ can be assumed to be constant for a given bias 
 at steady state.

The electrical power $W$ dissipated into the wires by Joule effect is then given by
\begin{eqnarray}
W= Ri^{2} = R i_{0}^{2}& & \left[ 1+ \frac{1}{2}\left(\frac{i_{1}}{i_{0}}\right)^{2} \left(1 +\cos{2\omega t}\right)\right. +\nonumber \\
&&+ 2 \left. \left(\frac{i_{1}}{i_{0}
}\right) \cos{\omega t} 
\right]
\label{eq:W}\end{eqnarray}
whose average value is
\begin{equation}
{\bar W}= W_{0}\left[ 1+ \frac{1}{2}\left(\frac{i_{1}}{i_{0}
}\right)^{2} \right]= W_{0}+ \mathcal{O}\left[\left(\frac{i_{1}}{i_{0}
}\right)^{2}\right]
\label{eq:barW}\end{equation}
with $W_{0}=R i_{0}^{2}.$  $\mathcal{O} \left[(\ldots)^{k}\right]$ means small terms of order $k$ or higher. In the worst case, the modulation amplitude contributes a few parts per thousands to the Joule effect. This confirms the assumption that the wire temperature is mainly determined by the d.c. bias.

The intensity $I$ of the emitted radiation is given by Stefan's law. At steady-state, 
the power input $W$ is balanced by heat losses due to the thermal conduction through  the wire supports and to the radiation emitted by the glowing wires. In this condition, it is intuitive to assume (and the validity of this assumption will be proved later on) that the temperature $T$ of the wires is proportional to $W,$ thus yielding $I\propto W^{4}.$ By expanding $W^{4}$ and keeping only leading order terms in $(i_{1}/i_{0})$, we get
\begin{eqnarray}
I(t)\propto  W_{0}
^{4}& & \left[  1 +14\left(\frac{i_{1}}{i_{0}}\right)^{2} \left(1+ \cos{2\omega t} \right)
+ \right.\nonumber\\
 &&
\left.
8\left(\frac{i_{1}}{i_{0}
}\right)\cos{\omega t} +
\right. \left.  \mathcal{O}\left(\frac{i_{1}}{i_{0}
}\right)^{3}
\right]
\label{eq:Ilight}\end{eqnarray}
The detector output is $\propto I$ and \eqnref{eq:Ilight} states that it contains modulation at the same frequency of the current modulation as well as at twice that frequency. 
The synchronous detection with the lock-in amplifier picks up only the amplitude of the $\cos{\omega t}$ term and its output $v$ is thus proportional to the current modulation amplitude
\begin{equation}
v= G W_{0}^{4} \left( \frac{i_{1}}{i_{0}}\right)=G W_{0}^{4} \left( \frac{V_{1}}{V_{0}}\right)
\label{eq:vli}\end{equation}
Here, $G$ is a suitable constant that includes the overall gain of the amplification chain, the detector sensitivity, the solid angle subtended by the wires at the detector, and so on.

As a check of the consistency of the previous approximations, the detector signal is monitored on a digital oscilloscope (Agilent, mod. DSO3102A) that also displays the signal spectrum. In all experimental conditions only the first harmonic is present, whereas second and higher order harmonics, if present, are below the noise level. This means that all terms of order $(i_{1}/i_{0})^{2}$ and higher are negligible and that the wire temperature is only set by $W_{0},$ as stated by \eqnref{eq:barW}.

\begin{figure}[b!]
\centering
\includegraphics{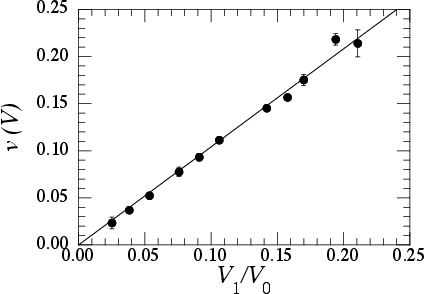}
\caption{\small Linear relationship between the lock-in output signal $v$ and the amplitude modulation ratio $V_1/V_0$ with $V_0 =6.6\,$V.\label{fig:Vlinear}}
\end{figure}

The linearity of the lock-in output as a function of the current modulation strength, \eqnref{eq:vli},
has been verified for quite a wide range of modulation amplitude, as reported in~\figref{fig:Vlinear}.
The amplitude of the a.c. modulation can be as high as 25\% of the d.c. bias $V_0$ and still the linearization procedure leading to \eqnref{eq:vli} retains its validity.

\begin{figure}[!t]
\centering
\includegraphics{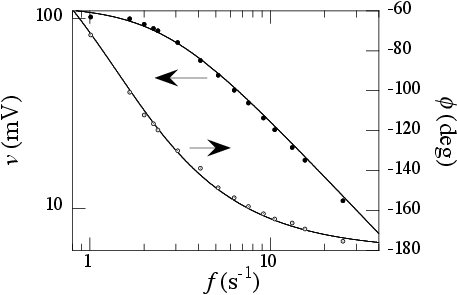}
\caption{\small Frequency response of the wires. Closed points: amplitude (left scale). Open points: phase $\phi$ (right scale). Lines: frequency response curves of a first-order low-pass filter. \label{fig:LP}} 
\end{figure}

The choice of the lock-in parameters depends on the frequency response of the wires that act as a natural low-pass filter because of  their thermal inertia. This behavior is shown in~\figref{fig:LP}, in which the amplitude of the rectified lock-in signal $v$ is reported as a function of the modulation frequency $f$ along with the signal phase $\phi.$ 
  The lines are the usual low-pass filter frequency response curves
  $v(f) ={v_0}/{\sqrt{1+(f/f_0)^2}}$  and $\phi(f)=-\left({360^\circ}/{\pi}\right)\tan^{-1}\left({f/f_0}\right).$
In the present case, $f_0\approx 2.5\,$s$^{-1}.$ In order to maximize the wires signal we have chosen $f\approx 1.7\,\mathrm{s}^{-1}$ as the working frequency.
 For the lock-in amplifier to correctly work, its integration time constant $\tau_{\mathrm{LI}}$ has to be set to a few seconds. If the polarizer were continuously rotated by energizing the d.c. motor with a constant voltage, the longest attainable rotation period would be $\approx 520\,$s. By so doing, the polarizer would rotate through a few degrees during the integration time of the lock-in amplifier. In order to get rid of the uncertainty thus introduced in the determination of the polarizer rotation angle $\theta$ and to improve the SNR, we have set $\tau_{\mathrm{LI}}=10\,$s and we have devised to rotate the polarizer through steps of finite amplitude (e.g., $5^\circ$). After each step, the polarizer is kept still for a time interval of  $60\,$s in order to allow the lock-in amplifier output to settle down. Once the settling interval has elapsed, the lock-in amplifier output is averaged over an interval $\tau=\tau_{\mathrm{LI}},$ and the whole procedure is repeated. 

\begin{figure}[b!]
\centering
\includegraphics{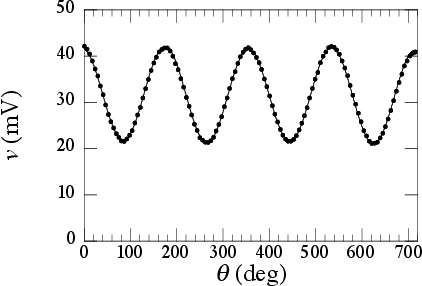}
\caption{\small Typical signal amplitude $v$ vs the polarizer rotation angle $\theta.$ Solid line:~\eqnref{eq:malusres}. Fitting parameters: $v_u=21.7 \, $mV, $v_p= 20.3\,$mV, and $v_r=0.12\,$ mV.
\label{fig:malusFit}}
\end{figure}

 A typical record of $v$ as a function of $\theta$ is shown in~\figref{fig:malusFit}.
According to Malus' law~\cite{jenkins}, 
the polarizer produces a $\cos^{2}{(\theta)}$ modulation of the light intensity at the detector, hence of the detector signal amplitude
\begin{equation}
v=v_u+ v_p \cos^{2}{(\theta-\theta_0)}
\label{eq:vpol}
\end{equation}
 where $\theta_0$ is the initial angle between polarizer and wire axis when the polarizer rotation is started and has no physical relevance. $v_u$ and $v_p$ are the amplitudes of the unpolarized and polarized components of the emitted light, respectively, and are determined by fitting~\eqnref{eq:vpol} to the data.

Actually, the experimental data also show a small $\cos\theta$ contribution that is caused by a tiny misalignment $\delta$ of the polarizer axis with respect to the optical axis of the system. Owing to the long length of the optical arm ($\approx 0.6 \,$m) and to the small detector area ($1\,$mm$^2$), the misalignement induces a periodic motion of the wires image across the detector itself, thereby leading to an intensity modulation with the same periodicity of the polarizer rotation. Therefore, data are fitted to \begin{equation}
v=v_u+ v_p \cos^{2}{(\theta-\theta_0)}+v_r\cos{\left(\theta-\theta_r\right)}
\label{eq:malusres}\end{equation}
where $v_r$ is the amplitude of the modulation due to misalignment and $\theta_r$ is the relative, irrelevant phase angle. It always occurs that the residual modulation amplitude $v_r$ is quite small with respect to $v_p.$ Typically, $v_r / v_p \ll 2\,\% ,$ compatible with $\delta\lesssim 1^\circ.$
Thus, the $v_r$ contribution does not alter the measured polarization value within the experimental accuracy. In \figref{fig:malusFit} the fitting curve~\eqnref{eq:malusres} is superimposed to the lock-in output, showing a very good agreement.

 The (average) polarization is computed as the polarization contrast
   \begin{equation}
 { \langle P\rangle}=\frac{v_p}{2v_u + v_p}
\label{eq:Pola}   
   \end{equation}
in which the factor of 2 extinction of the unpolarized light component due to the polarizer has been taken into account.

Typically, for each $i_0$ settings, i.e., for each $T,$ a long experimental run is carried on by recording at least 5 complete revolutions of the polarizer in order to improve the statistical accuracy of the experiment. During the whole run, the wire resistance is continuously recorded so as to check that the temperature remains constant within $1\,\%.$

\begin{figure}[!t]
\centering
\includegraphics{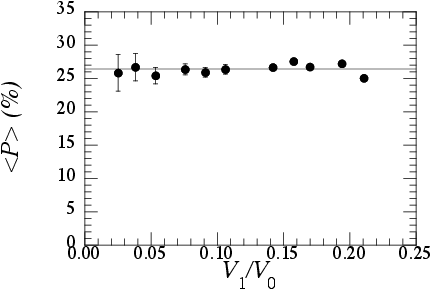} 
\caption{\small Average polarization $\langle P\rangle$ vs relative modulation width $V_1/V_0$ for $V_0=6.6\,$V.\label{fig:PindV1V0}}
\end{figure}

The lock-in output $v$ and the encoder output $\theta$ are fetched by a PC over a GPIB-IEEE 488 bus and are stored for offline processing. The data set of the long run is divided in subsets corresponding each  to one single polarizer turn. The data of each subset is fitted to~\eqnref{eq:malusres} by using well-known nonlinear least-sqlouares algorithms~\cite{bevington}. For each subset $j,$ the polarization $\langle{P}_j\rangle$ is computed with the aid of~\eqnref{eq:Pola}. Finally, the polarization for the long run  is obtained as the weighted average of the individual $\langle{P}_j\rangle$'s.
It is worth noting that this determination of $\langle P\rangle$ does not depend on the modulation amplitude, as shown
in~\figref{fig:PindV1V0}. This results confirms once more that the wire temperature is mainly set by the d.c. bias.

\subsection{Wire Temperature Determination}\label{sect:temperature}
\begin{figure}[b!]
\centering
\includegraphics{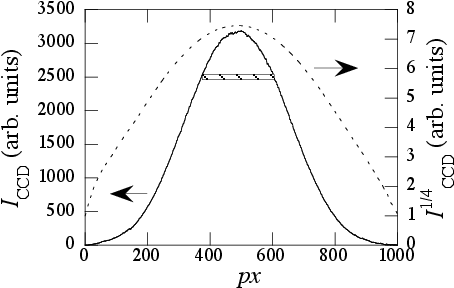}
\caption{\small Solid line: profile of the light intensity
$I_{\scriptscriptstyle\mathrm{CCD}}$ emitted by the wires as recorded by the
CCD camera (left scale). Dashed line: estimate of the temperature profile as
$I^{1/4}_{\scriptscriptstyle\mathrm{CCD}}$ (right scale). The coordinate along
the wires is measured in pixels ($1\, px$ = $4.65\, \mu$m). The wire image length is $\approx 1.5$ times shorter than true wire length because of the lens system magnification. The hatched box
represents the detector size in the wire image plane.\label{fig:CCDprofile}}
\end{figure}

In order to compare the experimental results with theory,
the polarization has to be connected to the wire temperature. Whereas
the determination of $\langle P\rangle $ is quite easy, the assessment of the
wire temperature is not straightforward.
 Even worse, $T$ is an ill-defined quantity. 
The wire ends are clamped on massive supports, which are in good
thermal contact with the room temperature parts of the apparatus that act as
the thermal equivalent of an electrical ground. In this situation, the
steady-state balance between the heat input by Joule heating and the heat loss
by thermal conduction through the supports and radiation emission leads to the
buildup of a non uniform temperature profile along the wires.

This intuitive expectation is confirmed by the analysis of the visible image of the wires produced by a CCD camera  (Lumenera, mod. Skynix2-1), whose output $I_{\scriptscriptstyle\mathrm{CCD}}$ is shown in~\figref{fig:CCDprofile} as a function of the
position $px$ in pixels along the wires. This picture has been recorded when the wires were glowing
yellow, viz., at an estimated central temperature $
T> 3000\,$K~\cite{wilkie2011}.

 The emission is maximum in the center and rapidly decreases towards the ends. A rough estimate of the temperature is obtained by applying Stefan's law as $T\propto I^{1/4}_{\scriptscriptstyle\mathrm{CCD}} $ and the corresponding temperature profile is also shown in~\figref{fig:CCDprofile}. It appears to be quite flat in the center where the
temperature, $T_{M},$ is much higher than the temperature $T_{0}$ at the boundaries.

\subsubsection{Integration of the energy balance equation\label{sect:enbal}} 

Actually, among several other practical reasons, the CCD camera cannot routinely be used 
in place of the MCT detector because its sensitivity is peaked in the visible region of the spectrum. 

The physical quantities that are directly measured in the experiment are the intensity $i_0$ of the d.c. current flowing in the wires, the d.c. voltage $V_0$ across them, and the intensity of the emitted radiation.
It is reasonable to expect that $T$ is directly related to the average electrical power $W_0=V_0 i_0$ dissipated in the wires. Thus, the comparison of the dependence of the emitted intensity $I$ and of the ohmic wire resistance $R$ on  $W_0$ with their temperature dependence computed by direct numerical integration of the energy balance equation allows us to establish the relationship between $W_0$ and $T.$

\begin{figure}[b!]
\centering  
\includegraphics{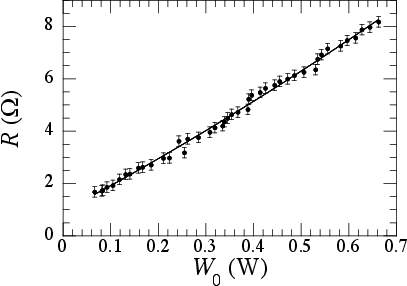}
\caption{\small $R$ vs $W_0$ for the $9\,\mu$m wires. Line: parabolic fit.\label{fig:4W9RvsW0}}
\end{figure}

In~\figref{fig:4W9RvsW0} we show the dependence of $R$ on $W_0$ for the $r=9\,\mu$m-wires.
The solid line is a parabolic fit. The Tungsten resistivity has a parabolic  temperature dependence~\cite{lide1995}
\begin{equation}
\rho (T)= \rho_0\left(1+\rho_1T +\rho_2 T^2\right)
\label{eq:rho}\end{equation}
in which $T$ is expressed in K, 
 $\rho_0= 48.0\,$n$\Omega\,$m, $\rho_1= 4.8297\times 10^{-3}\,$K$^{-1},$ and $\rho_2= 1.663\times 10^{-6}\,$K$^{-2}.$
Thus, we expect that a roughly linear $W_0-T$ relationship . 

In steady-state conditions, energy conservation requires that the energy losses due to thermal conduction and light emission equal the electrical power input, thereby leading to an ordinary differential equation (ODE) for $T.$
As the wire length is $L=7\,$mm and their radius is $r\ll L,$ we can approximate the bundle of the four wires located side by side with a single cylinder of the same length and cross sectional area $S=4\pi r^2.$ The effective surface area for emission is $S_l=4 (3/4)2\pi r L.$ The factor $3/4$ represents  the fraction of the lateral surface of the wires that emits toward the outer environment, whereas the radiation emitted by a fraction $1/4$ of the lateral surface is reabsorbed by nearby wires. The analysis of the $r=50\,\mu$m wire is simplified because only one wire is used.

Let us introduce a linear coordinate system $x$ along the cylinder whose ends are located in $x=0$ and $x=L$. Let  $K(T)$ and $e$ be the thermal conductivity and emissivity of tungsten, respectively. 
At steady state, $T$ does not vary with time, i.e., $\partial T/\partial t=0,$ and the energy balance equation reads
\begin{eqnarray}
\frac{\mathrm{d}^2 T}{\mathrm{d}x^2}
&+&\frac{\mathrm{d}\ln{K}}{\mathrm{d} T}\left( \frac{\mathrm{d}T}{\mathrm{d}x}\right)^2 +  \frac{\rho}{16 KS^2} i_0^2 - \nonumber \\
&-&\frac{S_l}{4KSL}\sigma e \left(T^4 -T_e^4\right)
=0
\label{eq:ODE}
\end{eqnarray}
$T_e\simeq 300\,$K is the environment temperature. $\sigma=5.67\times 10^{-8}\,$Wm$^{-2}$K$^{-4}$ is the Stefan-Boltzmann constant.
The nonlinear term $(\mathrm{d}T/\mathrm{d}x)^2$ arises from the dependence of $K$ on $T.$ 
According to literature data~\cite{lide1995} 
\begin{equation}
K(T)=
K_0 + K_1\mathrm{e}^{-T/T_K}
\label{eq:K}\end{equation}
in which $T$ is expressed in K. $K_0=96.07\,$W$\,$K$^{-1}$m$^{-1} ,$
$K_1=128.77\,$WK$^{-1}$m$^{-1}, $ and $T_K=555.03\,$K.  

\eqnref{eq:ODE} must be solved by enforcing the boundary conditions that, by symmetry, are $T(0)=T(L)=T_0.$ Unfortunately, $T_0$ is unknown because the thermal resistance of the wire supports is not known and $T_0$ must thus be treated as an adjustable parameter.

\eqnref{eq:ODE} can easily be cast in adimensional form amenable to numerical integration by scaling $x$ by $L$ and $T$ by $T_0.$ By letting $z=x/L$ and $y=T/T_0,$ we can write
\begin{eqnarray}
y^{\prime\prime} &+& \frac{\mathrm{d} \ln{K}}{\mathrm{d}y}{y^{\prime}}^2 +\left(\frac{L i_0}{4S}\right)^2\frac{\rho}{T_0 K} -
\nonumber \\
&-&\frac{3}{2} \frac{L^2 \sigma T_0^3}{K r }e\left(y^4 -y_e^4\right)=0
\label{eq:ODEad}
\end{eqnarray}
in which primes mean differentiation with respect to $z$ and 
$y_e =T_e/T_0.$ The boundary conditions now read $y(0)=y(1)=1.$

In order to numerically integrate \eqnref{eq:ODEad}, $T_0$ must be assigned. This is done in a self-consistent way by adjusting $T_0$ until the computed electrical resistance value 
\begin{equation}
R_c= \frac{L}{4S} \int\limits_0^1 \rho \left[ T_0 y(z) \right] \,\mathrm{d}z
\label{eq:erre}\end{equation}
matches the measured one, $R.$

Actually, the really big issue is the choice of the emissivity, which is highly problematic~\cite{touloukian}. Measurements show that it has quite a range, $0.02\lesssim e\lesssim 0.95$ and varies with sample,
temperature, surface preparation, atmosphere, purity, ageing, and, additionally, it is seen to change with wavelength and temperature.

We have therefore followed an empirical approach based on our experimental
observations. For the sake of conciseness we show here only the analysis for
the $r=9\,\mu$m-wires. The  same analysis has been carried out also for
the thicker wires leading to similar conclusions.

\subsubsection{Empirical determination of the emissivity}\label{sect:e}
\begin{figure}[t!]
\centering
\includegraphics{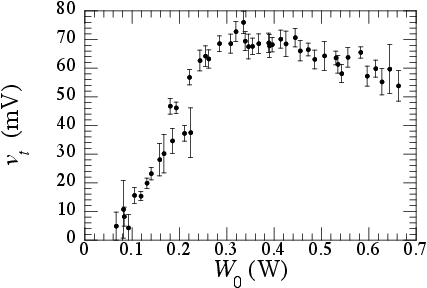}
\caption{\small Total intensity of the emitted light $v_t=2v_u+v_p$ vs $W_0$ for the $r=9\,\mu$m-wires. \label{fig:vtvsW0}}
\end{figure}

A first hint at how the emissivity $e$ might depend on $T$ comes from the analysis of the
dependence of total intensity  of the emitted light $v_t=2v_u+v_p$ on the
electrical power $W_0,$ as reported in \figref{fig:vtvsW0}.

$v_t$ increases with increasing Joule power up to $W_0\approx 0.3\,$W. Then, $v_t$ slightly decreases with a further increase of $W_0.$ This behavior is incompatible with the literature suggestion that $e$ might be roughly independent of $T.$  Actually, the detector output can be written as
\begin{eqnarray}
v_t &= & M \int\limits_0^L e\left[T(x)\right]\,\mathrm{d}x\times \nonumber\\ 
 & \times &
\left\{
 \int
\mathcal{P}\left(\lambda\right)t(\lambda) D(\lambda)B\left[\lambda, T(x) 
\right] \mathrm{d}\lambda
\right\}
\label{eq:detout}\end{eqnarray}
where we have assumed that $e$ does not vary too much with $\lambda$ in the
working wavelength range.
Here, $M$ is a factor that accounts for the effective radiating surface area,
for the solid angle of detector view, for the overall gain of the electronics,
and for all constants. $t(\lambda)$ is the transmission efficiency of the ZnSe window and $\mathcal{P}(\lambda)$
 is the spectral range of the polarizer and are both practically constant in the working wavelength range. 
$D(\lambda)$ is the detector responsivity that is given by a split Gaussian
function centered at $\lambda_0\approx 12.6\,\mu$m with left and right widths
$\sigma_{\pm}=(1.75, 0.66)\,\mu$m
\begin{equation}
\label{eq:detresp}
D(\lambda)\propto\left\{
\begin{array}{cc}
\exp{\left[-\left(\lambda-\lambda_{0}\right)^{2}/2\sigma_{+}^{2}\right]} & \mbox{for } \lambda\le \lambda_{0} \\ & \\
\exp{\left[-\left(\lambda-\lambda_{0}\right)^{2}/2\sigma_{-}^{2}\right]}
& \mbox{for }\lambda>\lambda_{0}
\end{array}
 \right. 
\end{equation}
$B(\lambda, T)$ is the Planck's blackbody radiation law
\begin{equation}
B(\lambda,T)\propto \frac{\lambda^{-5}}{\exp{\left(\lambda_T / \lambda\right)}-1}
\label{eq:Planck}\end{equation}
If \eqnref{eq:detresp} and \eqnref{eq:Planck} were used in \eqnref{eq:detout}  and if $e$ were assumed to be temperature independent, $v_{t}$ would turn out to be a monotonically increasing function of $T,$ in overt disagreement with the experimental result shown in \figref{fig:vtvsW0}. The behavior of $v_{t}$ can only be rationalized by assuming that $e$ is a decreasing function of $T$ and, hence, of $W_{0.}$

\begin{figure}[b!]
\centering  
\includegraphics{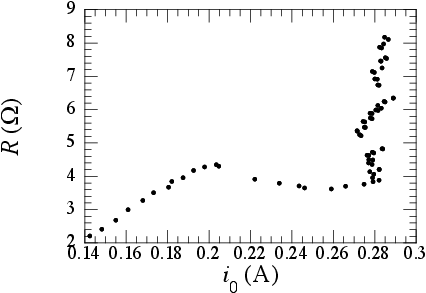}
\caption{\small $R$ vs $i_0$ for the $r= 9\,\mu$m-wires.\label{fig:4W9Rvsi0}}
\end{figure}

\begin{figure}[t!]
\centering  
\includegraphics{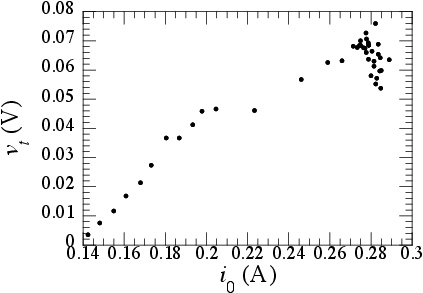}
\caption{\small $v_t$ vs $i_0$ for the $r= 9\,\mu$m-wires.\label{fig:4W9vtvsi0}}
\end{figure}

\begin{figure}[b!]
\centering  
\includegraphics{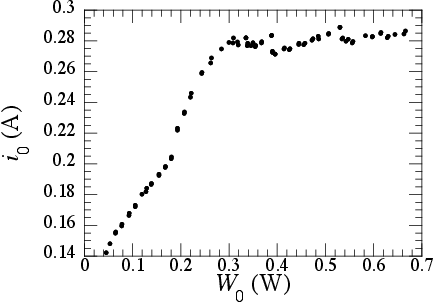}
\caption{\small $i_0$ vs $W_0$ for the $r= 9\,\mu$m-wires.\label{fig:4W9i0vsW0}}
\end{figure}

In order to proceed further, a delicate analysis of the data has to be carried out.
The data reported in \figref{fig:4W9RvsW0} and in \figref{fig:vtvsW0} can alternatively be plotted as a function of the d.c. current $i_0$ with astonishing results, as reported in \figref{fig:4W9Rvsi0} and \figref{fig:4W9vtvsi0}.
The measured resistance $R$ shown in \figref{fig:4W9Rvsi0} increases in an extremely rapid way in a very restricted d.c. current range as melting is approached. 
Also, the decrease of the detector output $v_t$ as a function of $W_0$ for $W_0\geq 0.3\,$W occurs in the same, very restricted  range of $i_0,$ as reported in \figref{fig:4W9vtvsi0}. 
This observation is supplemented by inspecting the behavior of $i_0$ vs $W_0$ reported in \figref{fig:4W9i0vsW0}. For $W_0\gtrsim0.3\,$W, $i_0$ does not vary very much although the Joule power keeps increasing.
We observe that the weird behavior of the measurements when they are plotted as a function of the current $i_{0}$ is not so unexpected because, although we experimentally set the d.c. bias voltage $V_0,$ it is the dissipated power $W_{0}$ that governs the physics of the problem by determining the wire temperature. Actually,
$i_{0}$ is only a derived quantity that depends on the wire resistance $R$ that is a function of the temperature, at which steady state is reached. 

It clearly appears that there is a very restricted range of d.c. current, over which the temperature rapidly increases. This fact cannot be attributed only to the increase of the Joule power but has to be traced back to a quite rapid decrease of the emissivity with increasing $T.$ In fact, if $e$ decreases, less energy is radiatively dissipated and $T$ increases faster than if $e$ were constant.

Wires break, as expected, when their temperature in the center, $T_M,$ reaches the melting value $T_m=3695\,$K~\cite{lide1995}. Within $1\,\%,$ melting occurred for $W_0 =W_m =0.675\,$W, corresponding to $i_0=0.285\,$A. For the $r=25\,\mu$m-wires melting occurred at $W_m\approx 4.68\, $W with $i_0\approx 2.08 \,$A and at $ W_m\approx 3.66\,$W with $i_0\approx 2.93\,$A for the $50\,\mu$m-wire. We recall that only $n_w=1$ wire is used for the largest diameter whereas $n_w=4 $ wires are mounted together for the other two diameters. 
\begin{figure}[t!]
\centering  
\includegraphics{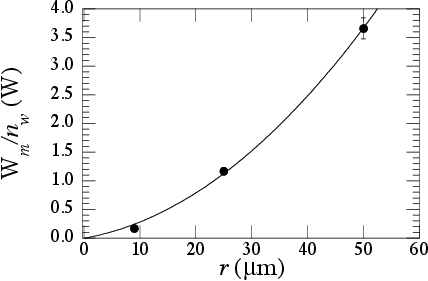}
\caption{\small $W_m/n_w$ vs $r$. Points: experiment. Solid line: Fit of type $W_m/n_w =a r^2+br,$ where $a$ and $b$ are fitting parameters.\label{fig:WmperNw}} 
\end{figure} 
In \figref{fig:WmperNw} we plot the power at melting per wire $W_m/n_w$ as a function of the wire radius $r.$ It turns out that $W_m/n_w$ is a zero-crossing parabola as a function of the wire radius $r.$
 This outcome has to be expected because both the heat capacity $C_p$ and the fusion enthalpy $H$ are proportional to the mass of the wires that is proportional to $r^2$ and the emitting surface $S$ is proportional to $r$ for wires of equal length.
 \begin{eqnarray}
  \frac{W_m}{n_w}& \propto & r^{2}\left\{ 
  \pi L \frac{\rho_{w}}{\mathcal{M}} \left[
  c_{p}\left(T_{m}-T_{e}
  \right) 
  \right]
  \right\} + 
   \nonumber\\ 
  & +&   r \left[
  2\pi L e\left( T_{m}^{4}-T_{e}^{4}
  \right)
  \right]
  \label{eq:EnthalpyofFusion}\end{eqnarray}
where $\rho_w,$ and $c_p$ are the mass density, the molar heat capacity, respectively, and $\mathcal{M}$ is the tungsten atomic weight. The contribution due to the fusion enthalpy can be safely neglected because of the tiny amount of wire (of the order of $10^{-5} $ to $10^{-4}\,$grams or less) that melts.
 \figref{fig:WmperNw} nicely confirms the validity of the experimental determination of the power at melting.
 
\begin{figure}[b!]
\centering  
\includegraphics{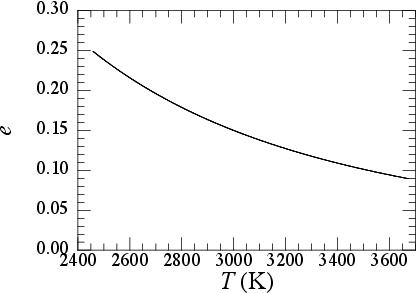}
\caption{\small Solution of \eqnref{eq:PJEQPRAD} $e$ vs $T$ evaluated for $i_0=0.285\,$A for the $r= 9\,\mu$m-wires.\label{fig:evsT}}
\end{figure}

By symmetry, \eqnref{eq:ODE} actually predicts that the temperature reaches its maximum $T_M$ in the wire center. An upper limit to $e$ can be set regardless of the value of the boundary temperature $T_0$ by equating the Joule power to the radiated power, i.e. by solving the equation
\begin{equation}
\frac{\rho(T)L}{4S}i_0^2=S_l\sigma e \left(T^4-T_e^4\right)
\label{eq:PJEQPRAD}
\end{equation}
with the current value $i_0\approx 0.285\,$A, at which melting occurred.
In this case, the wire temperature is constant all over the wire length $T(x)=T_M=T_0$ because \eqnref{eq:PJEQPRAD} yields 
\begin{equation}
\frac{\mathrm{d}^2 T}{\mathrm{d}x^2}+\frac{\mathrm{d}\ln{K}}{\mathrm{d}T}\left(\frac{\mathrm{d}T}{\mathrm{d}x}
\right)^2=0
\label{eq:Tunif}\end{equation} 
The solutions of \eqnref{eq:PJEQPRAD} in the restricted range, in which $i_0$ is nearly constant, are plotted in \figref{fig:evsT}.
The melting temperature can be achieved only if $e\lesssim 0.08\equiv e_M, $ provided that $T_0=T_M.$  Obviously, as it always is $T_0<T_M,$ $e_M$ is an upper limit to $e.$
Moreover, $0.277\,\mathrm{A}<i_0<0.285\,\mathrm{A}$ in the the range $0.3\,\mbox{W}\le W_0\le 0.675\,\mbox{W}.$ Hence, the curve in \figref{fig:evsT} can be assumed to roughly represent the behavior of $e$ as a function of $T.$ In the restricted temperature range shown in \figref{fig:evsT}, $e\propto T^{-\beta}$ with $2.2 < \beta < 2.5$ quite accurately fits the solution of \eqnref{eq:PJEQPRAD}.

In order to carry out in a  manageable way the numerical integration of \eqnref{eq:ODEad}, however, some tradeoffs have to be accepted, namely, $\beta $ has to be an integer. Moreover, we have to prevent $e$ getting larger than 1 at lower temperatures because it is unphysical. Thus, after some trial-and-error attempts, the following functional form for $e$ has been used
\begin{equation}
e\left( y, A, \alpha\right)=\frac{A}{1+(A-1)
y^\alpha}\qquad  (y=T/T_0)
\label{eq:eA}
\end{equation}
with $\alpha=2.$ $A$ is an adjustable parameter that has to be fixed by comparing the resistance value computed for $T_m$ with the measured one.

The functional form of $e$ in \eqnref{eq:eA} is such that $e$ is limited to 1 at the wire boundaries where $T=T_0.$ Even though this  value is too high, it has very small influence on the result because the region of the wires close to the boundaries emits very little compared to their central part according to Stefan's law.

In addition, as $T_M>T_0,$ \eqnref{eq:eA} gives $e\propto T^{-\alpha}$ near the center of the wires, as required by \eqnref{eq:PJEQPRAD}.
The value $\alpha=2$ is quite close to the $\beta$ value obtained by solving \eqnref{eq:PJEQPRAD} and allows the numerical integration of the ODE. By choosing  different values for $\alpha$, unphysical results are obtained. 
For instance, for $\alpha \geq 3,$ $T_0$ must be chosen unreasonably high to get agreement with the central wire temperature and numerical integration leads to the existence of wire regions characterized by the unacceptable property $T<T_{0}.$  Thus, the choice $\alpha =2$ is an reasonable compromise arising from a delicate balance between Joule dissipation, radiation, and thermal conduction.

The strategy of the ODE integration is to seek for the values of $A$ and $T_0$ that yield $T_M=T_m$ and $R_c=R_m\approx 8.18\,\Omega,$ the value of the resistance measured just at melting for $i_0=i_m=0.285\,$A.  

\begin{figure}[t!]
\centering  
\includegraphics{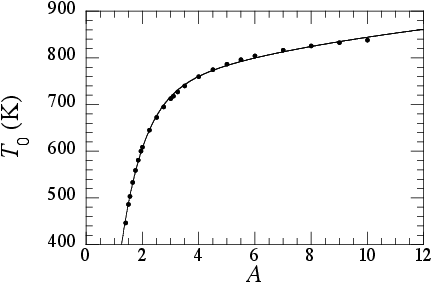}
\caption{\small  Relationship between the parameters $A$ and $T_0$ yielding $T_M=T_m$ upon integrating the ODE \eqnref{eq:ODEad} with $i_0=0.285\,$A for the $r= 9\,\mu$m-wires. The solid line is only an interpolating function. \label{fig:epsT0alfa2L7}}
\end{figure}

\begin{figure}[b!]
\centering  
\includegraphics{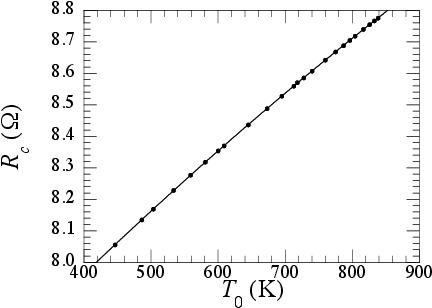}
\caption{\small $R_c$ vs $T_0$ with $T_0$ given in \figref{fig:epsT0alfa2L7}. The solid line is only an interpolating function. \label{fig:RcT0}}
\end{figure}

\begin{figure}[t!]
\centering  
\includegraphics{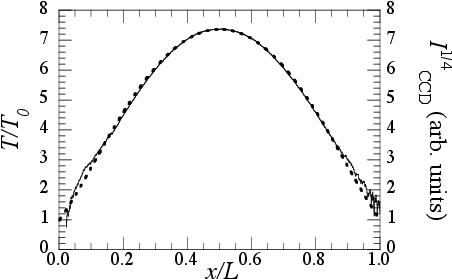}
\caption{\small The temperature profile at melting obtained by integrating the ODE \eqnref{eq:ODEad} (dashed line, left scale) is compared to the wire temperature profile measured with the CCD camera (solid line, right scale).\label{fig:TProfileatMelting}}
\end{figure}

In \figref{fig:epsT0alfa2L7} we show the relationship that the parameters $A$ and $T_0$ have to satisfy in order that the integration of the ODE \eqnref{eq:ODEad}  for $i_0=i_m$ yields a mid-wire temperature $T_M=T_m.$
At the same time, $R_c$ is computed according to \eqnref{eq:erre} and is plotted as a function of $T_0$ in \figref{fig:RcT0}.
By inspecting \figref{fig:epsT0alfa2L7} and \figref{fig:RcT0} we conclude that the required constraints, i.e., $T_M=T_m$ and $R_c=R_m$ are satisified by choosing the couple $(A\approx 1.535, \, T_0\approx 498\,\mbox{K}).$ The resulting temperature profile at melting is shown in \figref{fig:TProfileatMelting}
and is in very good agreement with the CCD camera measurements.

\begin{figure}[b!]
\centering  
\includegraphics{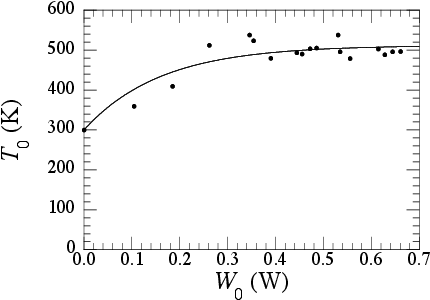}
\caption{\small  $W_0$-dependence of  $T_0$ that matches $R_{c}$ with $R.$ Line: fit curve. \label{fig:T0W0}}
\end{figure}

As a subsequent step, we have integrated the ODE \eqnref{eq:ODEad} by retaining the functional form \eqnref{eq:eA} while changing $T_0$ so as to achieve exact agreement between the computed resistance values $R_c$ with some selected measured resistance values $R,$ evenly distributed over the whole power range.
 By so doing we obtain the dependence of $T_0$ on $W_0$ reported in \figref{fig:T0W0}.
The solid line in the figure is only a fitting function 
of the form
\begin{equation}
    T_0 = a_{t} +b_{t} \left[
    1-\exp{\left(-W_0/c_{t}\right)}
    \right]\label{eq:T0fit}
\end{equation}
with $a_{t}=300\,$K, $b_{t}=212\,$K and $c_{t}=0.16\,$W and has no theoretical meaning. 

\begin{figure}[t!]
\centering   
\includegraphics{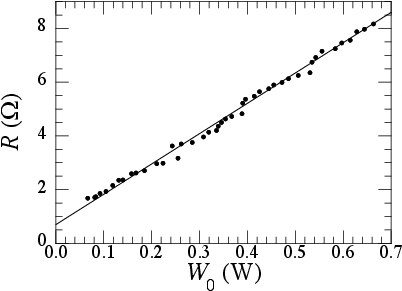}
\caption{\small Comparison of $R_c$ with $R$ as a function of $W_0 .$ Circles: experiment. Line: computation.
 \label{fig:RcR}}
\end{figure}

As a cross check of the validity of this approach, we integrated the ODE and computed $R_c$ for all experimental values of $i_0$ by using the fitting function~\eqnref{eq:T0fit} for $T_0.$ The comparison between $R_{c}$ (solid line) and $R$ (points) is shown in~\figref{fig:RcR}.  
The very good agreement between $R$ and $R_c$ lends credibility to our procedure. 

\subsubsection{Detector signal and wire temperature}
According to Stefan's law, each element of the wire of length $\mathrm{d}x$ contributes an amount $\propto T^4(x)\,\mathrm{d}x$ to the emitted light intensity.
The intensity of the light radiated from the central, hottest portion of the wires may even be a few thousands times larger than the contribution of the outer portions of the wires, as shown in \figref{fig:CCDprofile}.
\begin{figure}[b!]
\centering  
\includegraphics{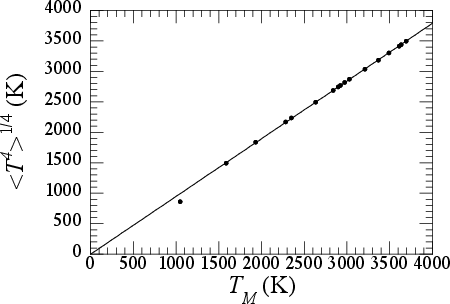}
\caption{\small $\langle T^4\rangle^{1/4}$ averaged over the central 40 \% portion of the wire vs $T_M.$ Line: equation $\langle T^4\rangle^{1/4}= 0.96T_M. $ 
 \label{fig:Tave4}}
\end{figure}

As previously explained, only the central portion of the wire is imaged onto the detector, whose output is thus proportional to a weighted average of $T^4$ over the central portion of the wire corresponding to $\approx 40\,\%$ of the total wire length.
We can define an effective radiation temperature $\bar T$ as
\begin{equation}
{\bar T} \equiv \langle T^4\rangle^{1/4}
 = \left[\frac{1}{x_f-x_i}\int\limits_{x_i}^{x_f} T^4(x)\mathrm{d}x\right]^{1/4}\label{eq:Te}\end{equation}
where $x_f - x_i$ is the length of the wire portion imaged onto the detector. $\bar T$ can be compared with the computed temperature in the center of the wire, $T_M$. It is found that $\bar T =0.96 T_M,$ as shown in~\figref{fig:Tave4}. Thus, with an error of a few \% at most, we can assign $T_M$ to the temperature $T$ of the detected radiation.

We finally plot the difference $T_M -T_e $ as a function of $ W_0.$
\begin{figure}[t!]
\centering  
\includegraphics{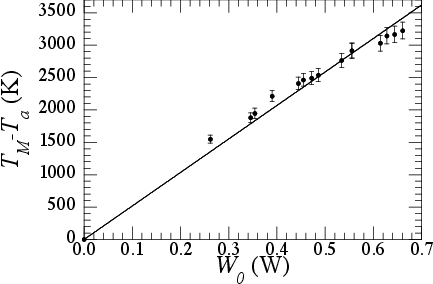}
\caption{\small  $T_M -T_e$ vs $W_0.$ The error bars are an estimate of the uncertainty on the computed values of $T_M.$ 
 \label{fig:096TMTa}}
\end{figure}
The data point in the origin is assumed on the basis that $T=T_e$ for $W_0=0.$ 
 The error bars are an estimate of the uncertainty of the computed values of $T_M$ due to the uncertainties on the determination on the several parameters used in the numerical integration of the ODE~\eqnref{eq:ODE}. The straight line in~\figref{fig:096TMTa} is a linear fit forced to pass through the origin. To a very good approximation, $T_M-T_e$ turns out to be a linear function of $W_0$ that can thus be cast in the form
\begin{equation}
  T\equiv T_M= T_e + \left(\frac{T_{m}-T_e}{W_{m}}\right)W_0
  \label{eq:TW}
\end{equation}
Here, $W_{m}$ is the electrical power at melting. \eqnref{eq:TW} is used to deduce the radiation temperature $T$ from the electrical power $W_0$ by using the values of $W_{m}$ measured for the wires of different diameters.

\subsection{Cross check of the validity of the temperature determination}\label{sect:crschk}
As a rough check of the validity of the linear relationship between $T $ and $W_0$ given in \eqnref{eq:TW} we have computed the emitted light intensity as a function of $T$ by using \eqnref{eq:detout}. We assume that the emissivity $e $ is given by the solution of \eqnref{eq:PJEQPRAD} in the region in which $i_0$ is roughly constant, i.e., for $W_0 \geq 0.3\,$W, whereas $e$ is assumed to be constant in the region for $W_0\leq 0.3\,$W, in which $i_0$ monotonically increases with increasing $W_0.$
\begin{figure}[b!]
\centering
\includegraphics{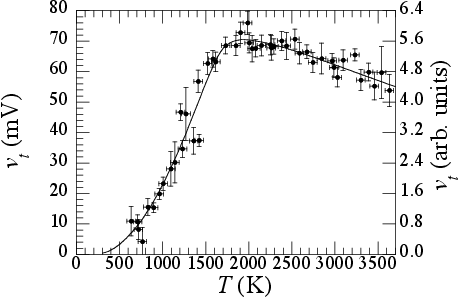}
\caption{\small $v_t$ vs $T.$ The data are the same plotted in~\figref{fig:vtvsW0}. \eqnref{eq:TW} has used to convert $W_0$ to $T.$ The solid line is~\eqnref{eq:detout}.
\label{fig:VtvsTepsTcross1700K}}
\end{figure}
The result of this calculation is compared with the experimental data in \figref{fig:VtvsTepsTcross1700K}.
We have to stress the fact that the solid line is computed as a function of $T$ whereas the experimental data are function of $W_0$ and are plotted as a function of $T $ by using the linear relationship \eqnref{eq:TW}. The agreement is very good and once again lends credibility to the adopted procedure.

 \subsection{Polarization Data and Comparison with Theory}\label{sect:data}

\begin{figure}[!t]
\centering
\includegraphics{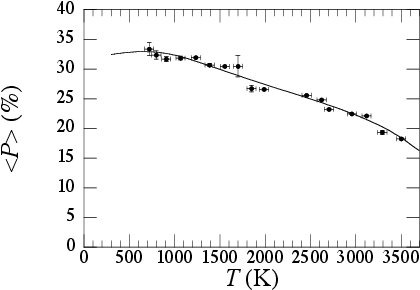} 
\caption{\small $\langle P\rangle $ vs $T$  for the $50-\mu$m wire.
Circles: experiment. Line: theory.
\label{fig:PvsTphi50}}  
\end{figure}

\begin{figure}[!b] 
\centering
\includegraphics{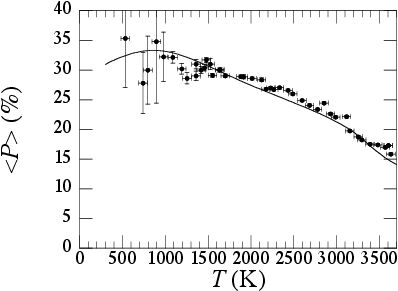}
\caption{\small $\langle P\rangle $ vs $T$  for the $25$-$\mu$m wires.
Circles: experiment. Line: theory.
\label{fig:PvsTphi25}} 
\end{figure}

\begin{figure}[!t]
\centering
\includegraphics{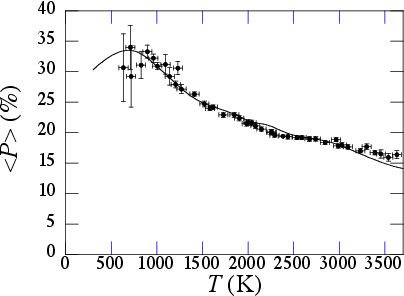} 
\caption{\small $\langle P\rangle $ vs $T$  for the $9$-$\mu$m wires.
Circles: experiment. Line: theory.
\label{fig:PvsTphi9}}  
\end{figure}

We are now able to present the polarization data as a function of the estimated wire temperature. In \figref{fig:PvsTphi50}, \figref{fig:PvsTphi25}, and \figref{fig:PvsTphi9} we show the experimental results for the $50$-, $25$-, and $9$-$\mu$m wires, respectively, along with the theoretical predictions.
$\langle P \rangle$ for all wire diameters strongly decreases with increasing $T.$ At low $T$ all data sets approach a polarization value of $\langle P\rangle \approx 35\,\%,$ whereas the polarization decreases towards $\langle P\rangle \approx 15\,\%$ near melting. At low $T$ the error bars are very large because the signal is tiny and the signal-to-noise ratio is unfavorable, whereas it greatly improves at higher $T.$

According to Kirchhoff's law~\cite{planck}, the absorptivity and emissivity of a body in thermodynamic equilibrium with the radiation field are equal. This conclusion has been proved true also if the body is freely radiating to the outside environment, provided that the local temperature of the body is well defined so that the energy distribution over the material states of the body is the equilibrium distribution~\cite{weinstein1960,burkhard1972,baltes1976}. 

Thus, the calculation of the wire emissivity proceeds via the calculation of the absorption efficiency (cross section per unit area) of a wire, which an electromagnetic wave is impinging on~\cite{hulst}. The wire is assumed to be a homogeneous circular cylinder of 
length $L$ and radius $r\ll L.$ The radiation scattered by the wire is observed at 
a distance $d$ from the wire in the plane crossing the wire at its
 midpoint and perpendicular to its axis. 
 
 The experimental conditions are such that 
 $r\sim 10^{-5}\,$m, $\lambda\sim 10^{-5}\,$m, $L\sim 10^{-2}\,$m,
 and $d\sim10^{-1}\,$m that yield the following inequalities: 
 $r^2/\lambda \sim 10^{-4} \,\mbox{m}\ll d\sim 10^{-1}\,\mbox{m}\ll L^2/\lambda\sim 10^{2}\,\mbox{m}.$ Under this conditions the scattered wave mainly has cylindrical symmetry. 
The scattered field is then obtained as the far-field solution for an infinitely
long circular cylinder~\cite{agdur1963}.  

The optical properties of the metal are 
described by a complex, $\lambda$ and $T$ dependent, relative permittivity $\epsilon\left(\lambda, T
\right).$ The cylinder surface is considered as a sharp boundary between the wire and 
the vacuum characterized by $\epsilon=1.$ As tungsten is a non magnetic material, its 
relative magnetic permeability is $\mu=1.$

Owing to the cylindrical symmetry of the problem, the electromagnetic field can
 be decomposed into transverse electric (TE)- and transverse magnetic (TM) modes. 
 TE modes are 
polarized with the electric field vector perpendicular
to the cylinder axis, whereas TM modes are 
polarized with the electric field vector parallel to the cylinder axis. For light of intensity 
$I_0$ incident perpendicularly upon a wire, the intensity of light scattered at an angle $\psi$
is  given by
\begin{equation}
I^{\dagger}(\psi)= {I_0} \left(\frac{2}{\pi k d}\right)\left\vert T^{\dagger}
(\psi)\right\vert^2\quad \dagger =\cases{\perp &\mbox{for TE modes}\cr
\parallel & \mbox{for TM modes}}
\label{eq:iscat}
\end{equation}
where $T^{\dagger}(\psi)$ is the scattering amplitude of the mode at hand and $k=2\pi/\lambda$ is the wave number {\em in vacuo}.

For TE modes, $T^{\perp}$ is given by
\begin{equation}
T^{\perp}(\psi)= a^{\perp}_0+ 2\sum\limits_{m=1}^{\infty}a^{\perp}_m \cos{\left(m\psi\right)}
\label{eq:tperp}
\end{equation}
in which the coefficients $a^\perp_m$ are obtained by enforcing the boundary condition that the 
electric field parallel to the cylinder axis vanishes at the wire surface, thus yielding
\begin{equation}
a^{\perp}_m = \frac{J^{\prime}_m (nkr) J_m (kr)-n J_m (nkr)J^{\prime}_m(kr)
}{{J^{\prime}_m (nkr)
H^{(2)}_m (kr) - n J_m (nkr)H^{(2)\prime}_m (kr)
}}
\label{eq:am}
\end{equation}
Here, $n=\sqrt{\epsilon}$ is the complex index of refraction of the material. $J_{m} $ are Bessel functions of the first kind and $H^{(2)}_{m}$ are Hankel functions of the second kind~\cite{arfken}. Primes indicate differentiation with respect to the argument.

Similarly, for TM modes, for which the component of the magnetic field parallel to the wire axis vanishes at the wire surface, $T^{\parallel}$ is given by
\begin{equation}
T^{\parallel}(\psi)=a^{\parallel}_{0}+2\sum\limits_{m=1}^{\infty}a^{\parallel}_{m}\cos{(m\psi)}
\label{eq:tpar}\end{equation}
with 
\begin{equation}
a^{\parallel}_{m}= \frac{nJ^{\prime}_{m}(nkr)J_{m}(kr) 
-J_{m}(nkr)J^{\prime}_{m}(kr)
}{nJ^{\prime}_{m}(nkr) H^{(2)}_{m}
(kr) - J_{m}(nkr) H^{(2)\prime}_{m} (kr)
}
\label{eq:bm}\end{equation}
The absorption efficiency factor $Q^{\dagger}_{\mathrm{abs}} $, i.e., the absorption cross section divided by the geometrical cross section of the wire, is given in terms of the extinction- and scattering efficiency factors $Q^{\dagger}_{\mathrm{ext}}$ and $Q^{\dagger}_{\mathrm{sca}},$ respectively, as $Q^{\dagger}_{\mathrm{abs}} =Q^{\dagger}_{\mathrm{ext}}-Q^{\dagger}_{\mathrm{sca}} .$
They are obtained as ($\dagger =\parallel, \, \perp$)
\begin{eqnarray}
  Q^{\dagger}_{\mathrm{ext}}&= &\frac{2}{kr}\mathtt{Re}\left(
 a^{\dagger}_{0} +2\sum\limits_{m=1}^{\infty}a^{\dagger}_{m}
 \right)  \label{eq:qdaggerext}\\ 
 Q^{\dagger}_{\mathrm{sca}} & =&\frac{2}{kr}\left(
 \left\vert a^{\dagger}_{0}\right\vert^{2} +2 \sum\limits_{m=1}^{\infty}\left\vert a^{\dagger}_{m}\right\vert^{2}
 \right)  \label{eq:qdaggersca}
 \end{eqnarray}
 The linear polarization of absorption is then defined~\cite{agdur1963} as
\begin{equation}
P_{\mathrm{abs}} =\frac{Q^{\perp}_{\mathrm{abs}} -Q^{\parallel}_{\mathrm{abs}}}{Q^{\perp}_{\mathrm{abs}} + Q^{\parallel}_{\mathrm{abs}}}=P\equiv P(\lambda ,T,r)
\label{eq:pol}\end{equation}
and, according to Kirchhoff's law, it is also the polarization $P$ of the light emitted by the wires.
The efficiency factors in~\eqnref{eq:qdaggerext} and~\eqnref{eq:qdaggersca} depend on  $\lambda$ and $T$ through the refraction index. The denominator of \eqnref{eq:pol} can be considered as the total intensity radiated by the wires, whereas the numerator is the net difference between the contributions of different polarization modes.

Actually, the polarization measurement, for a given $T,$ is an average over $\lambda$ of the light transmitted through the ZnSe window and IR polarizer and weighted by the detector responsivity $\mathcal{D}(\lambda)$. The polarizer can be approximated by a transmission coefficient of unity at maximum transmission and zero at minimum transmission in the present wavelength range and the ZnSe window transmission coefficient is nearly constant. 

The measured polarization of the light emitted by the wires is given by
\begin{equation}
\langle P  (T,r) \rangle = \frac{\langle Q^{\perp}_{\mathrm{abs}}\rangle  -\langle Q^{\parallel}_{\mathrm{abs}}\rangle }{\langle Q^{\perp}_{\mathrm{abs}}\rangle  + \langle Q^{\parallel}_{\mathrm{abs}}\rangle }
\label{eq:polave}\end{equation}
where the averages $\langle Q^{\dagger}_{\mathrm{abs}}\rangle$ are computed as
\begin{equation}
\langle Q^{\dagger}_{\mathrm{abs}}\rangle = \frac{1}{C}\int\mathcal{D}(\lambda) B(\lambda,T) Q^{\dagger}_{\mathrm{abs}}(\lambda,T,r) 
\,\mathrm{d}\lambda 
 \label{eq:aveQ}\end{equation} 
in which $B(\lambda, T)$ is the Planck's distribution,
$C=\int \mathcal{D}B\,\mathrm{d}\lambda $ is a normalization constant, and we have used the fact that $\mathcal{P}(\lambda) t(\lambda) $ is roughly constant. All integrals are carried out over the wavelength range, in which the product $B\mathcal{D}$ does not vanish.

In order to compute the efficiency factors as a function of $\lambda$ and $T,$ the refraction index $n(\lambda,T)=\sqrt{\epsilon(\lambda,T)}$ must be known. Unfortunately, the optical constants of tungsten have been measured or computed only in restricted wavelength- ($0.3\lesssim\lambda\lesssim 2.5\,\mu$m) and  temperature ($300<T<2400\,$K) ranges and the agreement between their different determinations is far from satisfactory~\cite{roberts1959,martin1965,barnes1966,Aksyutov:1977kx,Aksyutov:1980vn,ordal1983,rakic1998}.
On the contrary, our measurements greatly extend the investigated temperature range up to the melting point of tungsten $T_{m}\approx 3695\,$K and also extend the wavelength range because the HgCdTe detector is responsive to $\lambda $ up to $12\,\mu$m and more.

For computational purposes, an analytical expression for $\epsilon(\lambda, T) $ is required. Roberts~\cite{roberts1959} suggested a modified Drude-like expression for the relative permittivity
\begin{eqnarray}
\epsilon (\lambda,T)& =&1 + \sum\limits_{j=1}^{3}\frac{K_{0,j}(T)\lambda^{2}}{\lambda^{2} -\lambda^{2}_{s,j}(T)+ i \delta_{j}(T)\lambda_{s,j}(T)\lambda} - \nonumber \\
&& - \frac{\lambda^{2}}{2\pi c\epsilon_{0}}\sum\limits_{k=1}^{2}\frac{\sigma_{k}(T)}{\lambda_{r,k}(T)-i\lambda}
\label{eq:Drudeps}\end{eqnarray}
in which $\epsilon_{0}$ is the vacuum permittivity. The first sum represents the contribution of interband (or bound electrons) transitions, which are most important in the visible region, whereas the second sum gives the contribution of intraband (or free electron) transitions, which is dominant in the infrared region. 
The values of the coefficients $K_{0,j},$ $\lambda_{s,j},$ $\delta_{j}, $ $\sigma_{k},$ and $\lambda_{r,k} $ are tabulated for a few temperatures up to only $T=2400\,$ K. This formula agrees quite well with the results of another model for the optical properties of tungsten and other metals~\cite{rakic1998} and has been successfully used in the experiment aimed at measuring the wire polarization in the visible range~\cite{njp}. It has also been used for the theoretical computations of the heat radiation from long cylinders~\cite{Golyk:2012fk}.
For these reasons, we have used \eqnref{eq:Drudeps} also in the high temperature range because the coefficients are well behaved as a function of  $T$ and can be reasonably well extrapolated beyond the range given in literature~\cite{roberts1959}.

In \figref{fig:PvsTphi50} through \figref{fig:PvsTphi9} we compare the measured polarization with the results of the theoretical calculations.
In order to compute the averages \eqnref{eq:aveQ}, the lower integration limit is set to $\lambda_i\simeq 0.5\,\mu$m, below which the detector sensitivity vanishes. In order to set the upper integration limit, we note that the normalized distribution $B(\lambda,T) D(\lambda)/C$ is peaked at a wavelength $\lambda_m(T)$ that obeys a Wien-type relationship
\begin{equation}\label{eq:wienType}
    \lambda_m T^\gamma = F
\end{equation}
in which $\gamma\approx 1.167$ and $F\approx 12716\,$K$^\gamma\,$m
is a constant. For $\lambda>\lambda_m$ the distribution rapidly vanishes and the integration can be safely limited to $\lambda_f\approx \alpha \lambda_m,$ with $\alpha\approx 2.$

The agreement between experiment and theory is very satisfactory.
At all temperatures, even at the lowest ones, the values of the wire radii and dominant wavelengths are such that the experimental observations fall in the range of geometrical optics. Even for $T\sim 500\,$K, the lowest temperature at which a faint, though detectable signal can still be observed,  $\lambda_{\mathrm{max}}\approx 9\,\mu$m yields $k_\mathrm{min} r=2\pi r/\lambda_{\mathrm{max}} \approx 6 > 1$ for the thinnest wires and $k_\mathrm{min}r \approx 34 \gg 1$ for the thickest one. Moreover, in the near IR range both the real and imaginary part of the refraction index are of order $10$ or larger,  thus making $\vert nk_\mathrm{min}r\vert \gg 1.$

In the range of geometrical optics it is shown that $P_{\mathrm{abs}}\rightarrow 1/3 $ for large $kr$ and $nkr$~\cite{agdur1963}. Actually, the experimental data for all types of wires at low $T$ confirm the theoretical prediction. 
Further, the theory predicts that, in the range of geometrical optics, $P_{\mathrm{abs}}$ should decrease if $\vert n\vert$ decreases. Actually, the Drude-Roberts formula for the relative permittivity~\eqnref{eq:Drudeps}
predicts a decrease of $\vert n\vert$ with increasing $T,$ mainly due to the behavior of  $\mathtt{Im}\,(n).$ 
We can conclude that the overall decrease of the observed polarization is the result of the change of the Planck's distribution with $T.$ As $T$ is increased, it shifts to shorter $\lambda,$ for which the refraction index is smaller.

Once again, we have to point out that the optical properties of tungsten in the near infrared region for $\lambda\gtrsim 2.5\,\mu$m and for $T>2400\,$K are either unknown at all or affected by large uncertainties. 
Although the temperature dependence of the coefficients in the Drude-Roberts formula is quite well behaved, nobody guarantees {\em a priori} that their extrapolation at higher $T$ gives correct results. Nonetheless, the  satisfactory agreement of the computed polarization with the experimental data lends credibility to the extrapolation procedure.

  \subsection{Total Intensity Data and Comparison with Theory}\label{sect:dataET}
   The polarization is the ratio of the net difference of the radiation emitted by the two allowed modes to the total radiated intensity. The denominator of \eqnref{eq:pol} has to be interpreted as the total emitted intensity that can thus be computed from the knowledge of the efficiency factors as
    \begin{equation}
      I_t = \langle Q^\perp +Q^\parallel \rangle 
      \label{eq:ITOT}
  \end{equation}
 Experimentally, $I_t$ is proportional to the total amplitude of the lock-in signal
 $v_t = 2v_u + v_p ,$ as can be deduced from \eqnref{eq:vpol} and \eqnref{eq:Pola}.
 Thus, we obtain
 \begin{equation}
I_t = N v_t
\label{eq:ItVt}
 \end{equation}
 where $N$ is proportionality factor that accounts for the overall gain of the amplification chain. The l.h.s. of \eqnref{eq:ItVt} is theoretically computed and can thus be directly compared to the r.h.s. of the same equation, that is the quantity directly accessible to the experimenters.

\begin{figure}[!t]
\centering
\includegraphics{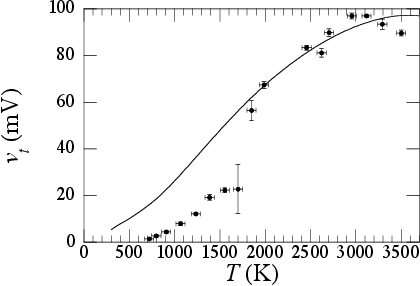}
\caption{\small $v_t $ vs $T$ for the $50$-$\mu$m wire.
Circles: experiment. Line: theory.
\label{fig:ItvsTphi50}}  
\end{figure}

\begin{figure}[!b]
\centering
\includegraphics{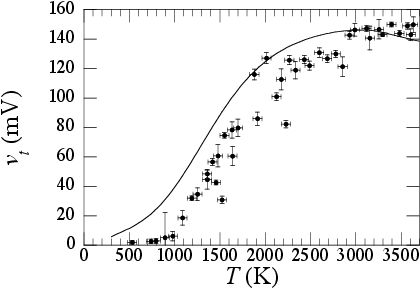}
\caption{\small $\langle P\rangle $ vs $T$  for the $25$-$\mu$m wires.
Circles: experiment. Line: theory.
\label{fig:ItvsTphi25}}  
\end{figure}

\begin{figure}[!t]
\centering
\includegraphics{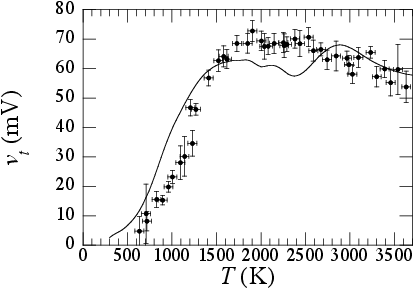}
\caption{\small $v_t $ vs $T$  for the $9$-$\mu$m wires.
Circles: experiment. Line: theory.
\label{fig:ItvsTphi9}}  
\end{figure}

In \figref{fig:ItvsTphi50}, \figref{fig:ItvsTphi25}, and \figref{fig:ItvsTphi9} we show the comparison between the measured and computed total intensity.
 The agreement between experiment and theory is fairly satisfactory, even though not as good as for the polarization, and leads to the conclusion that the determination of the temperature has been quite correct. 
The non perfect agreement of experiment and theory may be ascribed to several reasons. 
First of all, we recall that in the computation we have not made any special assumption about the functional form of the emissivity as we have previously done when comparing the experimental data to \eqnref{eq:detout}.

Secondly, we used wires supplied by different manufacturers so that the purity of the metals, hence their optical properties, might be slightly different. Moreover, at  high temperatures, ageing phenomena, which might act differently on wires of different diameter and properties, cannot be excluded. For instance, though the wires are being heated {\em in vacuo}, at very high $T$ the residual atmosphere could lead to the formation of WO$_{3}$ flocs.
 
In order to suggest one more possible reason for the worse agreement between theory and experiment for the total intensity, we note that $\langle P\rangle $ is a ratio of intensities and is thus less affected than $v_t$ by mid-term fluctuations, long-term aging, and similar effects.

We further note that, for the $9$-$\mu$m wires, the computed intensity near the maximum shows an oscillating behavior. We believe that this might due to an interference effect. Actually, for $T\approx 2400\,$K, $\lambda_m$ is such that $r/\lambda_m\approx 6$ is a small integer. For the wires of larger radius the same condition is only satisfied at much lower temperatures, where the experimental sensitivity is small. This assumption might also explain why the polarization of the light emitted by the $9$-$\mu$m wires shows a different concavity with respect to that emitted by the thicker ones as a function of $T.$

\section{Conclusions}\label{sect:conc} 

Thermal radiation has long been proved to be, at least, partially polarized. Spatial and temporal coherence of the light emitted by bodies of restricted geometry gives origin to phenomena, which are of interest for both fundamental physics and engineering applications.
We have measured the degree of linear polarization of thermal radiation emitted by thin, long tungsten wires in an extended temperature range up to the melting point. The measurements are carried out in a wavelength band across the infrared and visible region, in which the polarization is directed perpendicularly to the wires axis. We have observed a marked decrease of the polarization when $T$ is increased. 
We have been able to explain the temperature dependence of the polarization by extrapolating the validity of the Drude-type formula for the dielectric constant well beyond the temperature and wavelength ranges, for which it was originally proposed in literature.
Further measurements are now under way to investigate different metals in order to validate the present approach.
\section*{Acknowledgments}
We gratefully acknowledge stimulating discussions with Prof. G. Galeazzi, dr. G. Ruoso, G. Umbriaco, M. Guarise, and dr. G. Bimonte. We also thank dr. P.G. Antonini and Mr. E. Berto for technical assistance.

\end{document}